# The origins and concentrations of water, carbon, nitrogen and noble gases on Earth

Bernard Marty

CRPG-CNRS, Nancy-Université, 15 rue Notre Dame des Pauvres, 54500 Vandoeuvre-lès-Nancy, France

bmarty@crpg.cnrs-nancy.fr

ABSTRACT

The isotopic compositions of terrestrial hydrogen and nitrogen are clearly different from those of the nebular gas from which the solar system formed, and also differ from most of cometary values. Terrestrial N and H isotopic compositions are in the range of values characterizing primitive meteorites, which suggests that water, nitrogen, and other volatile elements on Earth originated from a cosmochemical reservoir that also sourced the parent bodies of primitive meteorites. Remnants of the proto-solar nebula (PSN) are still present in the mantle, presumably signing the sequestration of PSN gas at an early stage of planetary growth. The contribution of cometary volatiles appears limited to a few percents at most of the total volatile inventory of the Earth. The isotope signatures of H, N, Ne and Ar can be explained by mixing between two end-members of solar and chondritic compositions, respectively, and do not require isotopic fractionation during hydrodynamic escape of an early atmosphere.

The terrestrial inventory of $^{40}$Ar (produced by the decay of $^{40}$K throughout the Earth's history) suggests that a significant fraction of radiogenic argon may be still trapped in the silicate Earth. By normalizing other volatile element abundances to this isotope, it is proposed that the Earth is not as volatile-poor as previously thought. Our planet may indeed contain up to ~3000 ppm water (preferred range : 1000-3000 ppm), and up to ~500 ppm C, both largely sequestered in the solid Earth. This volatile content is equivalent to a ~2 (±1) % contribution of carbonaceous chondrite (CI-CM) material to a dry proto-Earth, which is higher than the contribution of chondritic material advocated to account for the platinum group element budget of the mantle. Such a (relatively) high contribution of volatile-rich matter is consistent with the accretion of a few wet planetesimals during Earth accretion, as proposed by recent dynamical models.

The abundance pattern of major volatile elements and of noble gases is also chondritic, with two notable exceptions. Nitrogen is depleted by one order of magnitude relative to water, carbon and most noble gases, which is consistent with either N retention in a mantle phase during magma generation, or trapping of N in the core. Xenon is also depleted by one order of magnitude, and enriched in heavy isotopes relative to chondritic or solar Xe (the so-called "xenon paradox"). This depletion and isotope fractionation might have taken place due to preferential ionization of xenon by UV light from the early Sun, either before Earth's formation on parent material, or during irradiation of the ancient atmosphere. The second possibility is consistent with a recent report of chondritic-like Xe in Archean sedimentary rocks that suggests that this process was still ongoing during the Archean eon (Pujol et al., 2011). If the depletion of Xe in the atmosphere was a long-term process that took place after the Earth-building events, then the amounts of atmospheric $^{129}$Xe and $^{131\text{-}136}$Xe, produced by the short-lived radioactivities of $^{129}$I ($T_{1/2}$ = 16 Ma) and $^{244}$Pu ($T_{1/2}$ = 82 Ma), respectively, need to be corrected for subsequent loss. Doing so, the I-Pu-Xe age of the Earth becomes ≤50 Ma after start of solar system formation, instead of ~120 Ma as computed with the present-day atmospheric Xe inventory.

## 1. Introduction

The origins and abundances of volatile elements (here, the highly volatile, also called atmophile, elements : water, carbon, nitrogen and noble gases) are still debated. Since the pioneering work of William W. Rubey in the 50's which showed that the oceans could not be derived from the alteration of rocks at the Earth's surface (Rubey, 1951), many studies have investigated the origin of terrestrial volatiles, often using noble gases as physical tracers (Albarède, 2009; Allègre et al., 1983; Caffee et al., 1999; Dauphas, 2003; Fisher, 1982; Holland et al., 2009; Javoy et al., 1986; Marty, 1989; Ozima, 1975; Pepin, 1991; Phinney et al., 1978; Porcelli et al., 2001; Staudacher and Allègre, 1982; Tolstikhin and Marty, 1998; Turner, 1989). Relevant data to constrain models are from the geological record (present-day mantle and atmosphere, ancient sedimentary rocks) and from the analysis of the composition of cosmochemical reservoirs : primitive and differentiated meteorites sampling the asteroid belt and inner planetary bodies, planetary atmospheres analyzed by space probes, and samples returned by dedicated space missions. In the first part of this contribution, available cosmochemical data are used to assess the origin of terrestrial volatiles.

Contrary to the case of refractory elements, it is not possible to estimate the volatile element content of the mantle source from the analysis of mantle-derived lavas because, owing to their low solubility in silicate melts, these elements have been lost and fractionated during magma degassing. Only remnants are found in mantle-derived magmas emitted along mid ocean ridge and oceanic intraplate volcanoes. Thus indirect approaches are necessary, that are based on the calibration of a volatile element of interest to a tracer whose geochemical cycle is well documented, e.g., $^3$He (Marty and Jambon, 1987), or Nb (Saal et al., 2002). In this contribution, the volatile (H, C, N, noble gases) compositions of the mantle regions sampled by mantle-derived lavas are revisited. The best documented reservoir is the depleted mantle (DM) sampled by mid-ocean ridge volcanism. Documenting directly the bulk mantle (BM) volatile content is not possible because the size, the location and the composition of the deep reservoir(s), presumably sampled by mantle plumes, are not known. A mass balance approach is attempted here based on the terrestrial budget of radiogenic $^{40}$Ar. From estimates of the K content of the Earth (Arevalo et al., 2009; Jochum et al., 1983), the amount of $^{40}$Ar produced by the decay of $^{40}$K ($T_{1/2}$ = 1.25 Ga) over the Earth's history can be computed. The amount of $^{40}$Ar still trapped in the silicate Earth is then obtained by subtracting the known amount of $^{40}$Ar present in the atmosphere (Allègre et al., 1996; Turner, 1989). The amount of other volatile species still trapped in the silicate Earth is then estimated based on their relations to $^{40}$Ar in mantle-derived lavas and gases.

## 2. Cosmochemical end-members

### 2.1. The chondrite connection

The isotopic composition of hydrogen, expressed as the D/H ratio, shows large variations among cosmochemical reservoirs (Fig. 1). The solar system formed from a cloud of gas and dust, the protosolar nebula (PSN), the composition of which being best preserved in the Sun and in the atmospheres of the giant planets. The elemental composition of the Sun is known from wavelength measurements of the solar photosphere and by comparison with primitive meteorites for refractory elements. The D/H ratio of the protosolar nebula, largely inferred from Jupiter's atmosphere measurements, has been estimated to be $25\pm5 \times 10^{-6}$ (Geiss and Gloecker, 1998) In contrast, the D/H ratio of chondrites is a factor 5 to 12 higher, with a distribution peak coinciding with the terrestrial (oceanic) D/H ratio of $156 \times 10^{-6}$ (e.g., Robert, 2003). Measurements of the D/H ratios of water in 5 comets (Bockelée-Morvan et al., 2004) show a ~two-fold enrichment in D compared to the chondritic signatures (Fig. 1). If such values characterize outer solar system objects, then the contribution of comets to terrestrial water must have been lower than ~10 % (Dauphas et al., 2000). Recently however, a terrestrial-like D/H ratio has been measured on comet 103P/Hartley 2, a Jupiter-family object (Hartogh et al., 2011), thus raising the possibility that terrestrial oceans could have originated from specific cometary material. The isotopic composition of nitrogen gives an independent constraint to this problem. The $^{15}$N/$^{14}$N ratio of the PSN has been recently documented thanks to the analysis of solar wind sampled by the Genesis mission spacecraft (Marty et al., 2011; Marty et al., 2010). The protosolar $^{15}$N/$^{14}$N ratio was $2.27\pm0.03 \times 10^{-3}$ or $\delta^{15}N = -383\pm8$ ‰ $\{\delta^{15}N = [(^{15}N/^{14}N)_{sample}/(^{15}N/^{14}N)_{standard} - 1] \times 1000$, where the standard $^{15}$N/$^{14}$N ratio is the composition of atmospheric $N_2$ (ATM) : $(^{15}N/^{14}N)_{ATM} = 3.678 \times 10^{-3}\}$. Not only the Earth but all inner solar system objects including Venus, the interior of Mars and meteorites are highly enriched in $^{15}$N relative to the PSN (Fig. 1). All comets measured so far (23 in total, including 103P/Hartley 2) are even richer in $^{15}$N than inner solar system objects as they present $^{15}$N/$^{14}$N ratios of HCN and/or CN within $6.8\pm0.3 \times 10^{-3}$ ($\delta^{15}N = +850$ ‰) (Arpigny et al., 2003; Bockelée-Morvan et al., 2004, 2008; Jehin et al., 2009). By combining D/H and $^{15}$N/$^{14}$N data, the genetic relationship between asteroidal material is re-enforced, and the difficulty of deriving terrestrial water from comets is substantiated by the high $^{15}N/^{14}N$ ratios of comets even for the one showing an ocean-like D/H composition (Fig. 1). However, there is a possible analytical caveat that will need to be resolved by further cometary analyses. The cometary $^{15}N/^{14}N$ ratios have been measured in CN and HCN, the only molecules that could be analyzed by remote sensing. Other cometary N-bearing molecules like $NH_3$ and $N_2$ (if di-nitrogen is effectively trapped in cometary ice, which is not granted) may not be isotopically equilibrated with CN nor HCN.

The isotopic composition of carbon shows much less pronounced variations among solar system reservoirs but are broadly consistent with those of H and N : solar carbon may be poor in the heavy, rare isotope $^{13}C$ compared to the Earth and meteorites. The $\delta^{13}C$ value of the solar wind from lunar soil measurements has been estimated to be around -100 ‰ relative to the PDB international standard (Hashizume et al., 2004), whereas most meteorites display $\delta^{13}C$ values close to the terrestrial value within ~20 ‰ (Kerridge, 1985).

The stable isotopes of hydrogen and nitrogen seem to indicate that the Earth and the parent bodies of meteorites acquired their signatures from a common ancestor reservoir that was isotopically well mixed. In detail, the average D/H and $^{15}N/^{14}N$ compositions of chondrites (excluding enstatite chondrites which do not contain significant amounts of water) are undistinguishable from the terrestrial surface inventory (Fig. 2), suggesting that the latter resulted from contributions of diverse materials, whose remnants are now found in the different meteoritic classes. The terrestrial mantle end-member is depleted in the heavy isotopes of hydrogen and nitrogen and points to the possible occurrence of a solar-like component at depth (Fig. 2) that is discussed in the next section.

*2.2. Solar-like gas in the mantle*

The best evidence for the occurrence of a solar component in the mantle comes from neon isotopes (Hiyagon et al., 1992; Honda et al., 1991; Marty, 1989; Sarda et al., 1988). Variations in the Ne isotopic composition in mantle-derived rocks and minerals indicate mixing between an atmospheric Ne component ($^{20}Ne/^{22}Ne$ = 9.80) and a mantle end-member with $^{20}Ne/^{22}Ne \geq 12.5$, likely to be solar in origin ($^{20}Ne/^{22}Ne$ = 13.8 for the solar wind, Grimberg et al., 2006; Meshik et al., 2007). The origin of this solar component is presently discussed between incorporation in Earth's accreting material of dust loaded with ancient solar wind (Raquin and Moreira, 2009; Trieloff et al., 2000), or direct trapping of a PSN component during Earth's formation (Yokochi and Marty, 2004). In support of the first possibility, most mantle-derived samples admit an upper limit of 12.5 for the $^{20}Ne/^{22}Ne$ ratios, a value similar to an end-member composition labelled Ne-B (Black, 1972) characterizing matter irradiated and processed by the solar corpuscular irradiation [e.g. lunar soils or (solar) irradiated meteorites]. Central to this possibility is the timing of planetary accretion. Trapping of solar ions onto accreting material requires the nebular gas to have been cleared off because otherwise the nebula would have been opaque to solar wind. If the nebular gas had already dissipated before the main stages of terrestrial accretion, this process could have been quantitatively efficient, especially because the corpuscular irradiation of the young Sun might have been enhanced by orders of magnitude. However, different lines of evidence suggest that the solar nebula lasted for several Ma after start of solar system formation (Podosek and Cassen, 1994), a time interval during which Mars-sized embryos were already grown up (Dauphas and Pourmand, 2011). If the proto-Earth reached such a significant size in a few Ma when the nebular gas was still present, a massive solar-like atmosphere could have been captured (Sasaki, 1990; Yokochi and Marty, 2004). Retention of a nebular, $H_2$-rich atmosphere might have taken place when the proto-Earth was larger than ~1000 km in diameter (Marty & Yokochi, 2006). The growing proto-Earth was likely to be partially or totally melted due to release of gravitational energy and of heat from the decays of short-lived nuclides, and the probable contribution of greenhouse warming by the massive atmosphere (Sasaki, 1990). Consequently, solar gases like neon would have been dissolved in the growing proto-Earth magmas. Marty and Yokochi (2006) estimated that the atmospheric pressure of a solar-like atmosphere around the proto-Earth necessary to provide the amount of solar Ne in the mantle (computed from $^{21}Ne$ produced by nuclear reactions induced to the decay of U in the mantle) was quite modest about $8 \times 10^3$ hPa (~8 bar). This value represents a lower limit since solar Ne could have been lost from the proto-Earth. In support of the nebular model, Yokochi and Marty (2004) have measured $^{20}Ne/^{22}Ne$ ratios higher than the Ne-B value of 12.5 and up to 13.0±0.2 in minerals from the Kola mantle plume, pointing to the occurrence of a remnant of nebular Ne in the mantle. Further progress on the origin of solar-like neon in the mantle will arise from future analyses of Ne isotopes in rocks and mineral related to mantle plumes.

The case of other noble gases is less straightforward. Mantle Ar, Kr, and Xe are dominated by a component having an atmospheric isotope composition, possibly gained by subduction of surface noble gases (Holland and Ballentine, 2006). However, the precise analysis of natural $CO_2$ gas of mantle origin suggests the occurrence of a minor non-atmospheric Xe component, either chondritic, or solar (Caffee et al., 1999; Holland and Ballentine, 2006) and of a chondritic Kr component (Holland et al., 2009). Interestingly, the Martian mantle, as sampled by the Chassigny meteorite, contains a solar Xe component (Mathew et al., 1998; Ott, 1988). Thus there is evidence not only from neon but also from krypton and xenon that the terrestrial mantle still contains a non atmospheric, cosmochemical component.

Mantle nitrogen is depleted in $^{15}N$ relative to the surface (Fig. 2), with $\delta^{15}N$ values estimated between -5‰ and -30‰ relative to ATM (Cartigny et al., 1998; Javoy et al., 1986; Marty and Zimmermann, 1999; Murty and Mohapatra, 1997). Among chondritic classes, enstatite chondrites are the only ones to show systematic depletions of $^{15}N$ relative to Earth, with $\delta^{15}N$ values of -30±10‰. This similarity lead Javoy et al. (1986) to propose that mantle nitrogen has kept a memory of its genetic relationship with enstatite chondrites (otherwise attested by similarities in key isotope ratios). Marty and Dauphas (2003) proposed alternatively that the negative $\delta^{15}N$ values of the mantle may represent past subduction of ancient surface N depleted in $^{15}N$. Because solar Ne is present in the mantle, the mantle $\delta^{15}N$ end-member value could also result from mixing about 8% solar N ($\delta^{15}N$ = -383‰) with surface N (0‰ for air, ~+5‰ for sediments). Estimates of the $^{22}Ne$ and $^{14}N$ contents of the present-day terrestrial mantle are $2 \times 10^{-14}$ mol/g and ~1 ppm, respectively, (Marty and Dauphas, 2003; Yokochi and Marty, 2004), yielding a $^{14}N/^{22}Ne$ molar ratio of ~$3\times10^6$, much higher than the solar ratio of ~12. Assuming that ~8% of mantle nitrogen was trapped from the PSN, the rest originating from a chondritic source from isotope mass balance), then solar N should have been enriched by a factor of ~$10^4$ relative to solar neon during trapping in the proto-Earth by dissolution of a PSN-like atmosphere into a magma ocean, or ~$10^5$ if the amount of nitrogen in the silicate Earth was higher initially as argued in section 4. Such an enrichment relative to inert neon could be the result of chemical trapping of N under reducing conditions. Indeed, the solubility of N in molten silicates increases significantly with decreasing oxygen fugacity (Libourel et al., 2003; Miyazaki et al., 2004). The oxygen fugacity during terrestrial differentiation has been estimated to be between 4 and 2 log units below the iron-wüstite buffer (Rubie et al., 2011; Wood et al., 2008). For IW-4, the solubility of N is increased by 3 orders of magnitude relative to that of noble gases (using the parameterized equation of Libourel et al. (2003), which is still lower than the requested enrichment factor of $10^4$-$10^5$. Thus the negative $\delta^{15}N$ values of the mantle may not trace the occurrence of solar-like N at depth but may instead represent a chondritic ancestor (Javoy et al., 1986) or past subduction of a fractionated N component (Marty and Dauphas, 2003).

Hydrogen in mantle-derived rocks and minerals presents a range of negative δD values (Fig. 2), down to -125‰ in magmatic minerals from the Hawaiian mantle plume (Deloule et al., 1991). It is not clear if this signature represents isotopic mass fractionation, for instance during subduction, or traces the occurrence of ~15% solar H (depleted in D) at depth, based on H isotope mass balance. Calculations involving Ne isotopes suggest independently that about 10 % water could have resulted from the dissolution of a primordial, solar-like atmosphere (Marty & Yokochi 2006), consistent with the D/H message. Further analyses of D/H ratios in mantle plume-related minerals may help resolving this other important issue.

*2.3. Hydrodynamic escape of an early atmosphere ?*

The difference in the isotopic composition of neon between the mantle and the atmosphere is commonly attributed to fractionation during hydrodynamic escape of an early atmosphere (e.g., Pepin, 2006, and refs. therein). However, isotope fractionation during escape of volatile species would also impact the stable isotope signatures of major volatiles like water, carbon, and nitrogen, which is not observed (Figs. 1 & 2). In fact, variations in the elemental and isotopic compositions of all terrestrial volatiles (but xenon, see section 4) can be accounted by mixing between two cosmochemical end-members and do not necessitate isotope discrimination during atmospheric escape. In a $^{20}Ne/^{22}Ne$ vs. $^{36}Ar/^{22}Ne$ diagram (Fig. 3), the atmospheric and the mantle values can be accounted for by mixing between a solar component and a chondritic component (Component A defined by Black, 1971, and Mazor et al., 1970, for carbonaceous chondrites). These end-member values do not fit an evolution following kinetic fractionation proportional to $m^{-1/2}$, as represented by the red curve in Fig. 3). These variations are consistent with the contribution of gas-rich bodies of carbonaceous chondrite compositions to a growing Earth that had already trapped a solar component presumably from the PSN. During impact degassing, chondritic noble gases would had been released into the growing terrestrial atmosphere and not readily trapped into the proto-mantle due to their chemical inertness. Furthermore, atmospheric argon has a $^{38}Ar/^{36}Ar$ ratio (0.188) in the range of chondritic values (Mazor et al., 1970; Ott, 2002), different from the solar ratio of 0.182, (Heber et al., 2009; Meshik et al., 2007). Thus there is no need to invoke isotopic fractionation for establishing the isotopic composition of light noble gases. The case of xenon, which is isotopically different from any potential cosmochemical end-member (solar or chondritic) compositions will be discussed in Section 4.

*2.4. Cometary shower during the Late Heavy Bombardment ?*

A cometary contribution to the terrestrial inventory of volatile elements appears limited from D/H (Dauphas et al., 2000; however, as noted above, this argument may weaken if further cometary D/H measurements confirm Earth-like compositions) and $^{15}N/^{14}N$ systematics (Jehin et al., 2009; Marty and Yokochi, 2006) The lunar surface bears scares of a late spike of bombardment around 3.8 Ga ago and this Late Heavy Bombardment (LHB) episode should have also affected the Earth by supplying an amount of ET material equivalent to a layer of 200 m on average on the Earth's surface. Gomes et al. (2005) proposed that the LHB consisted of ~50 % cometary material, due to the destabilisation of the Kuiper belt by catastrophic resonances of the giant planets. From mass balance consideration, Marty and Meibom (2007) have argued that the amount of noble gases in the atmosphere constrain the cometary contribution to less than 1 %, as comets are expected to be rich in noble gases (Bar-Nun and Owen, 1998). Measurement of cometary neon has been reported for a Stardust grain and indeed indicates that the original material was rich in Ne (Marty et al., 2008). From mass balance also, it can be shown that such a contribution would have limited effect on the isotopic compositions of H and N (Marty and Meibom, 2007). The overabundance of comets during the Earth/Moon bombardment in the LHB model of Gomes et al. (2005) is a well-known problem (A. Morbidelli, personal communication). It may suggest that most comets disintegrated during their journey into the inner solar system, before having a chance to collide with our planet and its satellite. Alternatively, a possible caveat is that the noble gas content of cometary matter has not yet been measured and is only estimated from experiments aimed at measuring the trapping efficiency of noble gases in growing amorphous ice (Owen et al., 1992). A test of the Gomes et al (2005)'s model and of a cometary origin for terrestrial oceans (Hartlog et al., 2011) will be to analyze cometary material, either in situ or in a pristine sample returned by a dedicated space mission, a high priority in the mind of the author.

**3. Reservoir inventories**

In this section, three different terrestrial reservoirs are considered : the atmosphere sensu lato (air, oceans, sediments, crust), the "depleted mantle" (DM) sourcing basalts at mid-ocean ridges, and the bulk mantle (BM) that represents the silicate Earth (excluding the atmosphere). The amounts of volatile elements in the different terrestrial reservoirs are normalized to the carbonaceous chondrite composition for comparison purpose (Figs. 4-8). Such normalisation is justified from stable isotopes, but is more speculative for elemental abundances. Carbon and nitrogen in meteorites are mostly hosted by a common phase which is insoluble organic matter (IOM), whereas a significant fraction of water (equivalent $OH^-$) is from aqueous alteration of the chondrite parent bodies. Different origins and alteration imply fractionation between elements. Fortunately, H and C correlate in un-metamorphosed carbonaceous chondrites, and the H/C ratio of the terrestrial mantle is within the range of values observed in chondrites (Hirschmann and Dasgupta, 2009). The case of the noble gases poses another problem. The noble gas isotopic compositions of chondrites are significantly different from those of the proto-solar nebula, being elementally enriched in heavy noble gases and isotopically enriched in heavy isotopes. Chondritic noble gases are trapped in very small weight fractions (< 1 % in mass) of the bulk meteorites (Lewis et al., 1975), the so-called Phase Q (the residue of acid -HF, HCl- attack of bulk chondrites) which is associated with IOM (Ott et al., 1981; Marrocchi et al., 2005). The abundance of noble gases varies from one meteorite to another, but roughly correlates with the C content, especially for carbonaceous chondrites (Otting and Zähringer, 1967). Despite variations in their absolute abundances, the noble gas abundance and isotope patterns are comparable among different meteorites, suggesting a common fractionation process from the protosolar nebula. Thus, adopting a chondritic reference pattern appears justified when discussing the origin(s) and fractionation of volatile elements during terrestrial accretion and subsequent evolution. In the following a mean carbonaceous chondrite (CC) composition is computed for normalization using data from Orgueil (CI; Mazor et al., 1970) and from Murchison (CM; Bogard et al., 1971) (Table 1), since these meteorites are among the most primitive ones and have suffered little thermal metamorphism.

*3.1. Earth's surface inventory*

Following Rubey (1951), the Earth's surface inventory includes the atmosphere, the oceans, and sedimentary rocks (and the continental and oceanic crusts, which are minor reservoirs compared to the other ones). Noble gases of the surface inventory are concentrated in the atmosphere (Ozima and Podosek, 2002). The water and carbon budget at the Earth's surface is from Hirschmann and Dasgupta (2009) and refs. therein, and that of nitrogen is from Marty and Dauphas (2003).

When normalized to CC (Fig. 4), the atmospheric abundance pattern is by no mean chondritic. The $H_2O/C$ ratio of the terrestrial surface is too high, as noted previously (Hirschmann and Dasgupta, 2009), and Ne, Ar and Kr are enriched by 1-2 orders of magnitude compared to major volatiles. This pattern cannot be ascribed to any known cosmochemical ancestor and has to be the result of planetary processing. The overabundance of noble gases could be a signature of a cometary contribution, possibly during the late heavy bombardment (Marty & Meibom, 2006) and/or the result of the depletion of major volatiles at the Earth's surface due to their trapping in the solid Earth, or a combination of both processes. To investigate such possibilities, it is necessary to evaluate the amount of volatile elements stored in the solid Earth.

*3.2. Depleted mantle (DM)*

The mantle source of mid-ocean ridge basalts (MORBs) is generally labelled depleted mantle (DM) in geochemical budgets. The term "depleted" refers to the most incompatible trace elements, like the light rare earth elements (REE), which are depleted in the mantle region source of MORBs as a consequence of past extraction of melts. MORBs depleted in light REE are labelled N-MORB ("N" for normal). Some MORBs are not depleted, and even enriched in light REE and other highly incompatible elements, and are termed E-MORBs ("E" for enriched; LeRoex, 1987). Although the distinction between these different mantle sources may be somewhat arbitrary (Albarède, 2005), the carbon/$^3$He ratio varies by a factor of ~4 between N- and E-MORB (Cartigny et al., 2001; Marty and Zimmermann, 1999; Nishio et al., 1999; Nishio et al., 1998), justifying to keep this nomenclature for volatile elements. The fraction of the different MORB types along mid ocean ridges is relatively well known, thus an average volatile composition of N-, and E-MORBs, weighted by their frequency of occurrence, results in an average MORB source equivalent to DM (Marty and Zimmermann, 1999).

The solubility of water is relatively high compared to other volatile elements, therefore water loss during degassing is minor compared to that of other volatiles. As a result, the abundance of $H_2O$ in the DM source is estimated from measurements of $H_2O$ contents in MORB glasses (which erupted at sufficient ocean depth to prevent significant water degassing). Global estimates converge to $150\pm50$ ppm $H_2O$ for the DM (Dixon et al., 2002; Michael, 1988; Saal et al., 2002; Salters and Stracke, 2004).

The carbon content of the MORB mantle can be estimated from the flux of volatile elements to the oceans. The flux of $^3$He is $1000\pm250$ mol/yr (Craig et al., 1975), which, for a magma generation rate at ridges of $20\pm4$ km$^3$/yr and a partial melting degree of $12\pm3$ % gives a $^3$He content of the MORB mantle of $2.0\pm0.8 \times 10^{-15}$ mol/g. The C/$^3$He ratio of the MORB mantle is $2.2\pm0.6 \times 10^9$ (Marty and Zimmermann, 1999; Resing et al., 2004), therefore (assuming a $^3$He flux of 1000 mol/yr) the content of the MORB mantle is $50\pm25$ ppm C. The fact that the C/$^3$He ratio of the N-MORB ($1.4\pm0.4 \times 10^9$) is significantly different from that of E-MORB ($5.9\pm1.3 \times 10^9$; Marty & Zimmermann, 1998) suggests that the C mantle content is heterogeneous, varying between $30\pm10$ ppm and $140\pm30$ ppm for the N-MORB and the E-MORB end-members, respectively.

Alternatively, C can be estimated from melt inclusions trapped in olivine in order to constrain the average mantle $CO_2$/Nb ratio, assuming that the C/Nb ratio is little fractionated and that the mantle Nb concentration is relatively well known. Saal et al. (2002) estimated this ratio to be ~250 and proposed a DM carbon content of $19\pm5$ ppm. Salters and Stracke (2004) used the same $CO_2$/Nb constraints to obtain a slightly different mantle Nb concentration, resulting in a mantle C concentration of $16\pm9$ ppm. Cartigny et al. (2008) proposed a different $CO_2$/Nb ratio (530) for the MORB mantle, which yields a higher C content of 44 ppm. It is possible that mantle heterogeneities could explain this discrepancy, but these estimates are within errors of the DM carbon concentration obtained by normalisation to $^3$He. Hirschmann and Dasgupta (2009) proposed lower C contents of the DM of $14\pm3$ ppm , probably because these estimates are more relevant to the N-MORB component of the DM . Here we select a value of $20\pm8$ ppm C for the DM, obtained from a weighted average of all estimates.

The nitrogen content of the DM mantle can be estimated in two ways. First, $N_2$ in MORB vesicles correlates well with radiogenic $^{40}$Ar*, yielding a near-constant N/$^{40}$Ar for this reservoir (Dauphas and Marty, 1999; Marty, 1995; Marty and Dauphas, 2003). From estimates of the $^3$He content of the mantle source of MORBs (see above), of the mean MORB $^3$He/$^4$He (8±1 Ra), and using a $^4$He/$^{40}$Ar* of 2.2 $^{+0.8}_{-0.6}$ (see below) for this reservoir, Marty and Dauphas (2003) estimated a N content of the MORB mantle source of 0.27±0.16 ppm. Independently, the DM nitrogen content can be also estimated from the C/N (molar) ratio of the MORB source. The resulting N content depends on the mantle source region considered. For the whole MORB mantle source (N- and E-MORBs), taking a C content of 50±25 ppm (see above), and a whole mantle C/N ratio of 535±224 (Marty and Zimmermann, 1999), one obtains a N content of 0.11±0.07 ppm. Taking the N-MORB mantle source as representative of the DM, the N content becomes 0.09±0.05 ppm (C = 20±8 ppm, see above; C/N = 270±100; Marty and Zimmermann, 1999). All these estimates compare well within errors and therefore it appears that there is not significant C/N fractionation during recycling. It is notable also that the N content obtained from C is lower by a factor of two than the one obtained from the noble gas calibration approach. This difference could be due to an overestimate of the $^3$He flux in the oceans (see above for discussion of C/$^3$He uncertainties). Alternatively, the C content of the DM mantle may have been underestimated, as suggested by Cartigny et al. (2001). Whatever the reason behind these differences, for the present exercise, a factor of 2 uncertainty is not crucial and is well within the uncertainties.

The noble gas abundance pattern is the mean composition of the popping rock (Moreira et al., 1998) and of the Bravo Dome gas (Ballentine et al., 2005). The absolute amounts were anchored using the C/$^3$He ratio and the DM carbon content.

The DM abundance pattern is lower than the range of compositions that would have been carried by a late veener consistent with the abundance of platinum group elements in the mantle (Fig. 4). However, the DM represents only a fraction of the mantle inventory, which is estimated in the next subsection.

*3.3. Bulk mantle (BM)*

The bulk mantle (BM) is defined as the silicate Earth, excluding the surface inventory. For non specialists of mantle geochemistry, it may look strange that this reservoir is not equivalent to the one (DM) discussed previously which sources the largest flux of mantle-derived magmas. This is because all budgets of incompatible elements indicate that the DM does not contain enough of these elements to fit the terrestrial composition and that there must be other reservoir(s) that host a large fraction of these. These reservoirs are not readily sampled by magmas reaching the surface, and their surface expressions may be volcanic provinces related to mantle plumes. Even in those cases, it is not clear which deep reservoirs are sampled. Furthermore, their sizes is model-dependent. Thus derivation of the BM abundance of the volatile elements should not depend on the reservoir sizes and is estimated here from a global mass balance, which involves the $^{40}Ar$ budget of the Earth. As outlined in the Introduction, this budget indicates that only a fraction of $^{40}Ar$ produced by the decay of $^{40}K$ is now in the atmosphere, the rest presumably being in the silicate Earth. The most recent estimate of the terrestrial K content for the bulk silicate Earth (BSE) is 280±60 ppm (1 σ, Arevalo et al., 2009), so that $^{40}K$ has produced 3.9±0.9 x $10^{18}$ moles of $^{40}Ar$ over 4.5 Ga. 1.6 x $10^{18}$ moles of $^{40}Ar$ are in the atmosphere (Ozima and Podosek, 2002), and Arevalo et al. (2009) estimated the maximum $^{40}Ar$ amount in the continents to be 3.5 x $10^{17}$ mol. Thus the mantle and continental crust may still contain 2.3±0.8 x $10^{18}$ mol $^{40}Ar$. This mass balance implies that ~50 % of radiogenic $^{40}Ar$ is trapped at depth (Allègre et al., 1996; Turner, 1989). The K content of the Earth is, however, subject to discussion. For instance, Jochum et al. (1983) proposed a silicate Earth K/U ratio of ~12,000 that results in a K content of 250 ppm (still consistent with the Arevalo et al's value given the uncertainty of 60 ppm for the latter). In a more extreme view, Albarède (1998) remarked that the K content could be off by a factor of two, suggesting the possibility that almost all $^{40}Ar$ produced by $^{40}K$ is now in the atmosphere. However, recent geoneutrino measurements (Gando et al., 2011) are consistent with a high U(+Th) content of the Earth, and therefore a similarly high K content, although uncertainties are still too high to permit a quantitative assessment of these concentrations.

The $N/^{40}Ar$ ratio of mantle volatiles shows limited variation in either MORBs or plume-derived magmas, and Marty and Dauphas (2003) estimated that the silicate Earth $N/^{40}Ar$ ratio is 160±40. Hence, the nitrogen content of the silicate Earth (excluding the atmosphere) is 3.6±1.8 x $10^{20}$ moles (0.84±0.43 ppm for a silicate Earth mass of $4x10^{27}$ g). Note that this mass balance does not make any assumption about the size of mantle reservoirs since the $N/^{40}Ar$ ratios are comparable among the surface, the DM and the BM inventories.

The volatile content of the bulk mantle can be derived from calibration to either $^{40}Ar$ or N. The flux of mid-ocean ridge basalts dominates magma production from the mantle. The C/N ratio of all MORB types has been estimated at 535±224 (Marty and Zimmermann, 1999). This ratio represents primarily the DM reservoir and the value corresponding to mantle plume should also be considered if C and N are fractionated during mantle extraction. C/N data for lavas associated with mantle plume provinces are scarce, but available data (submarine lava with high $^3$He/$^4$He ratios from the Red Sea, Hawaii and the North Fidji Basin, lavas from the Pacific hot spot area like the society islands) are within the MORB range (Marty and Zimmermann, 1999). The corresponding C content is 1.9±1.3 x 10$^{23}$ moles (580±380 ppm for the silicate Earth).

The carbon content of the mantle can also be computed from C/$^4$He ratios of plume-related gases for which there are much more data (Réunion, Hawaii, Iceland ; Craig et al., 1978; Hilton et al., 1998; Marty et al., 1991; Sedwick et al., 1994)  and of rock/mineral fluid inclusions from plume provinces (Trull et al., 1993). The log mean ratio of hotspot-associated fluids and fluid inclusions is 1.05±0.32 x 10$^5$ (n = 52). For these samples, $^{40}$Ar data are not generally available (not measured or air-contaminated for volcanic and geothermal gases), so instead I use the $^4$He/$^{40}$Ar production/accumulation ratio in order to calculate the source region C concentration. This approach is valid because C/He ratios do not vary significantly during magmatic or mantle processes (Fischer et al., 2009; Resing et al., 2004). In Fig. 5 the mantle source $^4$He/$^{40}$Ar production/accumulation ratio is modelled using published values of the mantle K/U ratio (Arevalo et al., 2009; Vlastelic et al., 2006).  The residence time of incompatible elements in the MORB mantle has been estimated to be between 0.6 and 1.0 Ga (Galer and Onions, 1985), thus for such time durations, the $^4$He/$^{40}$Ar ratios are within $2.2^{+0.8}_{-0.6}$. The bulk mantle accumulation time admits an upper limit of 4.5 Ga, therefore, despite the lower K/U ratio of OIBs, it is likely the OIB source reservoir has a $^4$He/$^{40}$Ar of 1.9 ±0.5, very similar to that of MORBs.

With the bulk mantle $^{40}$Ar concentrations calculated above, the corresponding BSE carbon content is 3.8±2.7 x 10$^{23}$ moles (1164±544 ppm), comparable to the above estimate but with a larger uncertainty. From these two approaches, I use a weighted average C amount of 2.5±1.0 x 10$^{23}$ moles (765±300 ppm) for carbon sequestrated in the interior of the silicate Earth, equivalent to 530±210 ppm C (Table 1) for the bulk Earth when normalized to the total mass of our planet (5.98 x 10$^{27}$ g). The water amount still in the silicate Earth is computed with the estimate of the mean mantle H/C mass ratio by (Hirschmann and Dasgupta, 2009), which yields a water content of 2700±1350 ppm for the bulk Earth (Table 1).

As shown by Trieloff and Kunz (2005), the noble gas abundance patterns of plume-related magmas are not significantly different from those of MORB's and I use the same abundance pattern as for the MORB source. The absolute abundances are anchored in two ways. First, as stated above, the mantle $^{40}Ar$ content is reasonably known. For the bulk mantle, a $^{40}Ar/^{36}Ar$ ratio of 6,000±2000 appears representative (Trieloff and Kunz, 2005), which yields a $^{36}Ar$ abundance of 8 x $10^{-14}$ mol/g. The $^{40}Ar/^{36}Ar$ ratios are notably derived from correlations between $^{40}Ar/^{36}Ar$ and $^{20}Ne/^{22}Ne$ ratios and extrapolated to the solar value for the latter (Marty et al., 1998), so they are unlikely to be perturbed by shallow atmospheric contamination of the magmas. Second, the C/$^{3}He$ ratio of both plumes and MORBs are not significantly different (Trull et al., 1993) and average 3±1 x $10^{9}$, giving a comparable $^{36}Ar$ content of 13 x $10^{-14}$ mol/g. The abundances of volatile elements in terrestrial reservoirs are listed in Table 1 together with their uncertainties.

*3.4. Comparison between Atmosphere and Bulk Mantle abundance*

The major volatile/noble gas patterns of the atmosphere and the bulk mantle reservoirs (Fig. 6) are somewhat symmetrical : the major, chemically reactive volatile elements are mostly stored in the mantle, whereas the inert noble gases are largely in the atmosphere. This fractionation could be due to either preferential degassing of the chemically inert noble gases, with carbon, water and nitrogen being sequestrated in the mantle during early degassing. This possibility requires more reducing conditions of the ancient mantle than the present day one to retain these elements (Kadik et al., 2011). Alternatively, the fractionation may instead indicate preferential recycling of major volatiles into the mantle, with much more limited recycling of the noble gases. Notably, the BM inventories for water and carbon are high: up to 90 % of water may be in the mantle, and up to 97 % carbon may be stored at depth. If these mass balance budgets are correct, then the amounts of $H_2O$ and C at the surface might have been buffered by the mantle inventories, and could have decreased with time if recycling was responsible for the fractionation between major volatiles and noble gases. Such massive recycling should have occurred early in the geological history in order to get mild environmental conditions at the Earth's surface and allow the existence of liquid water as early as the Hadean eon (Sleep et al., 2001).

**4. The volatile content of the Earth**

The bulk Earth volatile content is obtained by adding the ATM and BM inventories (Fig. 7). Overall, a chondritic abundance for volatile elements emerges, consistent with the evidence of a genetic link with chondrites from H and N isotopes. The bulk Earth abundance of noble gases is dominated by the atmospheric inventory, and Ne and Ar isotopic ratios are effectively within the chondritic values. The isotopic composition of atmospheric krypton is slightly different from the chondritic composition, which could be the result of atmospheric escape to a lesser extent than Xe (see below). The case of solar-like Ne in the mantle requires fine tuning : solar-like neon is indeed present in the mantle, but its entire terrestrial budget is dominated by a component more akin to chondrites (Fig. 3).

The volatile abundances estimated here are high with respect to those of other works (Bolfan-Casanova, 2005; Dixon et al., 2002; Hirschmann and Dasgupta, 2009). Uncertainties on the BM mantle abundances include those on the K abundance and therefore the results are fairly robust as far as orders of magnitude are concerned. The bulk Earth water content between of 2700±1350 ppm, corresponding to 10±5 ocean masses, is one order of magnitude higher than estimates of the upper mantle water content based on the $H_2O/Ce$ ratios ( 50-200 ppm, ≤ 1 ocean mass in the mantle, Michael, 1988; Saal et al., 2002). It is, however, consistent with a bulk Earth abundance of 1300 ppm (range : 550-1900 ppm) obtained using the measured $H_2O/K_2O$ ratios in MORBs and OIBs (Jambon and Zimmermann, 1990). The lower limit of the present estimate is also marginally consistent with the range of values for the mantle source of plumes (300-1000 ppm; Dixon et al., 2002). The present budget assumes that radiogenic $^{40}Ar$ in the atmosphere did not escape to space (if this would have been the case, the computed mantle content would be lower since it is computed by removing the atmospheric budget from the total amount of $^{40}Ar$ produced from $^{40}K$). This possibility is unlikely however, since it is not supported by identical $^{38}Ar/^{36}Ar$ ratios in the atmosphere and in chondrites. Another source of uncertainty is the potassium budget of the Earth. Indeed the 280 ppm K content of the Earth given by Arevalo et al. (2009) may be at the upper range of possible values (F. Albarède, personal communication). Taking instead the 250 ppm K estimate derived from the canonical U/K ratio of Jochum et al. (1983) would reduce the bulk Earth water content to 1700 ppm, closer to independent estimates of the bulk Earth water content. In view of these uncertainties, a range of 1000-3000 ppm for the water content of the bulk Earth is compatible with the range given by Jambon and Zimmermann (1990) and Dixon et al. (2002), and is preferred here.

The volatile abundances presented here correspond to the contribution of ~2 (±1) % of CC material (Table 1). This fractional contribution seems to be also seen in the abundance of terrestrial halogens: terrestrial abundances of Cl, Br and I are also consistent with a contribution of 2±1 % of carbonaceous chondrite material (Fig. 8). The abundance of F would requires, however, a much higher contribution of 17 %, for a reason unknown to the author.

The oxygen (Javoy et al., 1986), chromium (Trinquier et al., 2007), titanium (Leya et al., 2008) and molybdenum (Dauphas et al., 2002) isotope signatures of the Earth point to a enstatite chondrite (EC) composition rather than to a CC ancestor. However, EC are very reduced and do not contain significant amounts of water. Hence the classical model of heterogeneous accretion of Earth starting with a dry, reducing material and evolving towards more oxidizing contributions by addition of different material (Javoy et al., 1986; Wänke and Dreibus, 1988) appears substantiated by the present inventory. The contribution of volatile-rich bodies from the snow line region is in fact a logical consequence of planetary growth in the inner planet region : the locally dry bodies that were present in the Earth's forming region became depleted due to their accretion, which allowed more distant wet planetesimals to contribute to the proto-Earth as accretion proceeded (Morbidelli et al., 2000; Raymond et al., 2009; Walsh et al., 2011).

*The case of Nitrogen*

There are two notable exceptions to the chondritic abundance pattern : nitrogen and xenon are both depleted by ~one order of magnitude relative to other volatile elements. For N, this depletion can be due to trapping in the interior of the Earth. Under mantle oxygen fugacity conditions, nitrogen is presumably incompatible and presents a low solubility in basaltic melt similar to that of noble gases (Libourel et al., 2003; Miyazaki et al., 2004; Roskosz et al., 2006). However this might have not be the case in the early, more reducing mantle conditions (Kadik et al., 2011), and/or in the presence of water (Roskosz et al., 2006). Furthermore, preliminary metal-silicate partition experiments suggest that nitrogen is siderophile, with a partition coefficient of 1 to 10 in favour of iron at high pressure (Bouhifd et al., 2010; Kadik et al., 2011). Hence the "missing" N could have been sequestrated in the core and this possibility is consistent with the accretion of wet planetesimals during terrestrial differentiation, since their inventory is higher than the PGE one. This implies that carbon and hydrogen were not quantitatively partitioned in the core.

*The case of Xenon*

Xenon is depleted by one order of magnitude relative to other volatile elements when normalized to the chondritic composition (Fig. 7). Furthermore, atmospheric xenon is enriched in the heavy isotopes by 3-4 % amu$^{-1}$ relative to chondritic and solar compositions. In contrast, the isotopic composition of atmospheric krypton is little fractionated (< 1 % amu$^{-1}$) relative to that of cosmochemical end-members. This discrepancy, known as the xenon paradox, has led to sophisticated models of atmospheric evolution coupled with mantle geodynamics (Pepin, 1991; Tolstikhin and Marty, 1998) and cometary contributions (Dauphas, 2003; Owen et al., 1992) that could explain terrestrial noble gas patterns under ad hoc conditions, but which fail to account for the fact that atmospheric Martian volatiles present a comparable paradox. Because the other volatile elements, not only noble gases but also major volatiles, are in chondritic proportions and have isotopic compositions close to Chondritic (or solar for Ne), xenon might have been initially chondritic too, and could have been affected by a process that enriched the heavy isotopes and depleted its abundance relative to other volatiles. It has been proposed that xenon could be sequestrated in minerals at high pressure (Sanloup et al., 2005), but this possibility cannot account for the specific Xe isotope composition of the atmosphere. Processes able to fractionate the isotopes of Xe at the percent level require ionization of xenon (Bernatowicz and Fahey, 1986; Frick et al., 1979; Hohenberg et al., 2002; Marrocchi et al., 2011; Ponganis et al., 1997). Such processing should have affected mostly xenon and not the other noble gases (although there is room for fractionating to a much lesser extent Kr). It may be related to the specific electronic structure of Xe, which makes Xe the most reactive element among noble gases. Xenon has the lowest ionization potential among noble gases, $CO_2$, $CH_4$, $N_2$, and $H_2O$. In experiments aimed at fractionating Xe isotopes upon ionization, ionized Xe implanted in solids is enriched the heavy isotopes, whereas neutral Xe is much less efficiently trapped and is not isotopically fractionated. It is possible that Xe was fractionated and depleted due to interactions with UV light from the young Sun in the gas/dust medium in the inner solar system before planetesimal formation, since the irradiation flux was orders of magnitude higher than Today's (Ribas et al., 2011). Ionized Xe might have escaped via the magnetic field of the early Sun and only a small fraction of Xe ions implanted into dust and enriched in heavy isotopes would have been preserved. This possibility is, however, not consistent with the fact that none of the meteorites documented so far show atmospheric-like Xe.

Recently, xenon having an isotopic composition intermediate between the atmospheric and the chondritic ones has been documented in Archean ( ≤ 3Ga-old) sedimentary rocks (Fig. 9), suggesting that the atmosphere at that time had a less evolved Xe isotopic composition (Pujol et al., 2011). Thus Xe depletion and isotope fractionation could have taken place by progressive escape of Xe ions from the atmosphere, possibly through interaction with UV light from the young Sun. Assuming a Rayleigh type isotope evolution for atmospheric Xe :

$$(^{i+1}Xe/^{i}Xe)_{Present}/(^{i+1}Xe/^{i}Xe)_{Initial} = f^{(\alpha-1)}$$

where f is the depletion factor (= 1/20; Pepin, 2006), the instantaneous fractionation factor $\alpha$ requires enrichment of 1.3 % in heavy isotopes for Xe remaining in the atmosphere. This fractionation is clearly within the range of values observed for Xe trapped in solids during Xe ionization experiments. For instance, Marrocchi et al. (2011) observed an enrichment of 1.4 % per amu for Xe trapped in growing carbon films under high frequency ionization of a rarefied Xe atmosphere. In the Hadean/Archean atmosphere, ionized Xe could have been partly trapped and enriched in the heavy isotopes in atmospheric dust while other Xe isotopes would have escaped. Alternatively, ionized $Xe^+$ could have been pushed up by escaping $H^+$ ions, with the heavier Xe isotopes remaining preferentially (K. Zahnle, personal communication). In both cases however remains the problem of keeping Xe ionized long enough to permit its escape from the atmosphere.

The Xe isotopes produced by the extinct radioactivities of $^{129}$I and $^{244}$Pu permit to compute closure ages for the atmosphere and, by inference, for the Earth. $^{129}$I was present at the birth of the Solar system and decayed with a half-life of 16 Ma into $^{129}$Xe. Mono-isotopic excesses of $^{129}$Xe are found both in the mantle and in the atmosphere. Because most terrestrial Xe is within the atmosphere (Fig. 6), it is straightforward to compute a $^{129}$I-$^{129}$Xe retention age of about 100 Ma (see for instance (Pepin and Porcelli, 2002) for data and computation) from the terrestrial budget of I (Deruelle et al., 1992) and the amount of radiogenic $^{129}$Xe in the atmosphere. A comparable age of ~120 Ma is obtained by combining the $^{129}$I-$^{129}$Xe and $^{244}$Pu-$^{136}$Xe ($T_{1/2}$ = 82 Ma) systems (Ozima and Podosek, 1999; Pepin and Porcelli, 2002; Yokochi and Marty, 2005). These ages are consistent with the Pb-Pb model age of the Earth (Albarède, 2009) and with the identical W isotope composition of the Moon and Earth (Touboul et al., 2007). If xenon was lost long after Earth's formation, then the ~20 fold Xe depletion in the atmosphere might have occurred when production of Xe isotopes by extinct radioactivities was achieved. For instance, Xe trapped in Archean rocks 3.5 Ga ago presents an isotopic composition intermediate between Chondritic and modern Atmospheric, indicating in the model of Pujol et al. (2011) that only half of initial Xe was already lost at that epoch. Thus the atmospheric contents of Xe isotopes produced by the decays of extinct radioactivities need to be corrected for subsequent loss and multiplied by a factor of ~20 to get the real age. Doing so, one finds a I-Pu-Xe closure age of 40-50 Ma for the Earth instead of ~120 Ma. This younger age is fully consistent with the Hf-W age of the Earth relative to primitive meteorites (Yin et al., 2002), and does not contradict the identity of tungsten isotopic ratios between Earth and Moon if these two bodies have similar Hf/W ratios (Touboul et al., 2007).

## 5. Conclusions

The origin of terrestrial volatile elements has been re-evaluated in the light of the stable and noble gas isotopes. These tracers point to a chondritic, rather than solar or cometary, source. A small ($\leq 10$ %) fraction of mantle volatiles might have been derived from the protosolar nebula during an early stage of the proto-Earth growth. Light noble gas compositions are consistent with mixing between chondritic and solar end-members, rather than due to fractionation of an originally solar component in the atmosphere during hydrodynamic escape. Likewise, the stable isotopes of H and N do not support atmospheric escape because their compositions are very similar to those of primitive meteorites.

A bulk Earth inventory has been achieved by calibration to noble gas isotopes produced by extant radioactivities. Most of water and carbon are within the Earth whereas noble gases are mostly in the atmosphere. This fractionation testifies for active exchanges of volatile elements between the mantle and the surface. The high water and carbon contents of the bulk Earth (equivalent to the contribution of 2±1% of carbonaceous chondrite) suggest the accretion of volatile elements from wet planetesimals during the main Earth forming events, rather than contribution of a late veener after Earth's differentiation required from the platinum group element budget.

Bulk Earth water, carbon, neon, argon and krypton are in chondritic relative proportions, but nitrogen and xenon are depleted by one order of magnitude relative to the other volatiles.

Nitrogen could be partitioned in the core or in a retentive mantle phase. The depletion of xenon could be the result of Xe preferential escape during the geological history of the Earth, as suggested by the occurrence of chondritic-like Xe in Archean sediments, presumably representing the composition of the ancient atmosphere (Pujol et al., 2011). If this model is correct, then the $^{129}$I-$^{129}$Xe age of the Earth needs to be corrected for Xe loss that occurred after Earth's formation, which gives an age of ~40-50 Ma after start of solar system formation rather than ~120 Ma as previously inferred.

**Acknowledgments**

This work was supported by the European Research Council under the European Community's Seventh Framework Programme (FP7/2007-2013 grant agreement no. 267255 to B.M.). I am grateful to Pete Burnard for having contributed the $^{4}$He/$^{40}$Ar subsection and for helpful comments on the ms. Although the ideas presented here are on my own responsibility, this work benefitted from discussions with Francis Albarède, Nicolas Dauphas, Colin Goldblatt, Allessandro Morbidelli, Robert Pepin, Magali Pujol, Kevin Zahnle, among others, from constructive reviews by A. Morbidelli and F. Albarède, and from careful editing by Yannick Ricard. This is CRPG contribution n° 2135.

| | Solar mol/g | CC mol/g | +/- | ATM mol/g | DM mol/g | +/- | BM mol/g | +/- | Bulk Earth ppm | +/- | Bulk Earth/ CC |
|---|---|---|---|---|---|---|---|---|---|---|---|
| $^{12}$C | 8.60E-11 | 2.94E-03 | 5.31E-04 | 1.29E-06 | 1.67E-06 | 6.67E-07 | 6.38E-05 | 2.5E-05 | 526 | 206 | 1.49% |
| $^{14}$N | 1.78E-11 | 1.09E-04 | 1.63E-05 | 5.98E-08 | 6.43E-09 | 3.57E-09 | 9.0E-08 | 4.6E-08 | 1.68 | 0.85 | 0.11% |
| H$_2$O | 1.17E-10 | 6.60E-03 | 2.10E-03 | 1.48E-05 | 8.33E-06 | 2.78E-06 | 2.0E-04 | 9.6E-05 | 2699 | 1347 | 2.27% |
| $^{22}$Ne | 2.13E-12 | 1.62E-12 | 4.33E-13 | 2.27E-14 | 3.42E-16 | 1.71E-16 | 5.8E-15 | 3.2E-15 | | | 1.64% |
| $^{36}$Ar | 6.31E-13 | 3.47E-11 | 2.78E-12 | 9.58E-13 | 2.72E-15 | 1.88E-15 | 7.8E-14 | 4.3E-14 | | | 2.91% |
| $^{84}$Kr | 3.15E-16 | 4.25E-13 | 1.79E-14 | 1.98E-14 | 1.51E-16 | 7.57E-17 | 1.9E-15 | 1.1E-15 | | | 4.95% |
| $^{130}$Xe | 7.88E-18 | 5.38E-14 | 1.46E-14 | 1.08E-16 | 2.84E-18 | 1.95E-18 | 2.6E-17 | 6.8E-18 | | | 0.23% |

**Table 1 : Reservoir compositions**

The solar abundances are anchored to a $^{22}$Ne content equal to the average of carbonaceous chondrites (from Mazor et al., 1970), the abundances of the other elements are computed with data from Anders and Grevesse (1989). The chondritic composition (CI-CM) is an average of Orgueil and Murchison data. Sources of data; carbon and nitrogen : Kerridge (1985), H$_2$O Robert (2003); noble gases : Mazor et al. (1970), Bogard et al. (1971). The atmospheric inventory comprises the atmosphere sensu stricto, the oceans and sedimentary rocks. H$_2$O : Lecuyer et al. (1998), carbon : Hirschmann and Dasgupta (2009), nitrogen : Marty and Dauphas (2003), noble gases : Ozima and Podosek (2002). The atmospheric abundances are normalized to the mass of the Earth (5.98 x 10$^{27}$ g) for comparison purpose. "DM" is the mantle region sampled by mid-ocean ridges (see text for data sources and explanations on the computations). "Bulk mantle" are concentrations in the silicate Earth, not comprising the surface inventory. Both DM and BM abundances are normalized to the mass of the silicate Earth (4 x 10$^{27}$ g). Their computations based on the $^{40}$Ar amount of the silicate Earth and, for C, also on the C/He ratios of mantle gases, are described in the text. "Bulk Earth" are the global concentrations obtained by adding the atmospheric and bulk mantle quantities and normalized to the total mass of the Earth. Uncertainties are 1 σ.

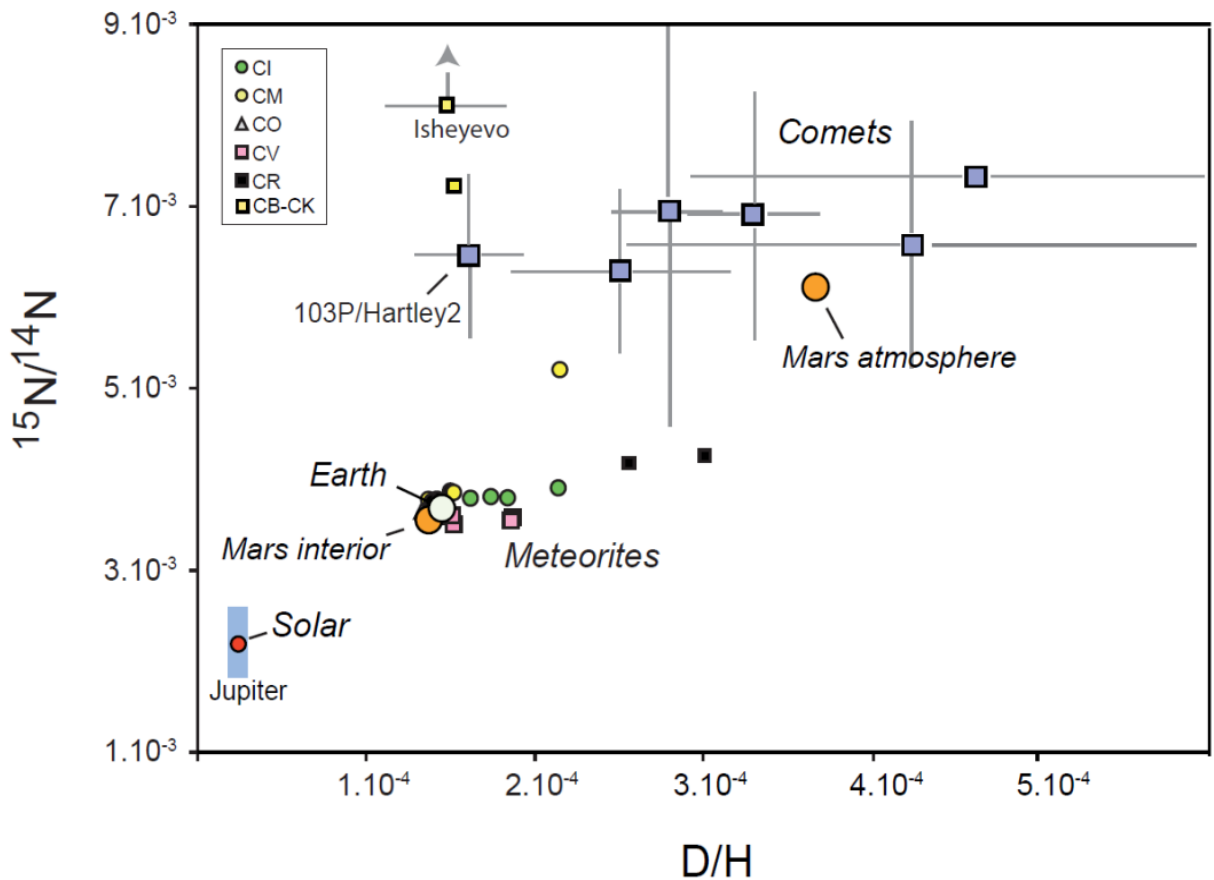

**Fig 1** : Co-variations of D/H and $^{15}N/^{14}N$ ratios among solar system reservoirs and objects. Meteorite, solar and Jupiter data are from refs in Marty et al., (2011), cometary data are from Hartogh et al. (2011), Jehin et al. (2009), Manfroid et al. (2009) Meech et al. (2011), Mars data are from Deloule (2002), Leshin et al. (1996), Murty and Mohapatra (1997), Owen et al. (1977). CI; CM, CO, CV, CR and CB-CK refer to the different carbonaceous chondrite groups. As their name imply, carbonaceous are rich in carbon and other volatile elements compared to ordinary chondrites. CI and CM are presumably the most primitive carbonaceous chondrite groups. The composition of the Martian atmosphere is attributed to isotopic fractionation of N and H during escape. However, the fact that the Martian atmosphere composition is within the field of cometary values could indicate a cometary contribution that could be more prominent on Mars than on Earth due to the small size of the Martian atmosphere. The newly determined D/H value of comet 103P/Hartley 2 is terrestrial-like reviving the possibility of a cometary origin for oceans, but its $^{15}N/^{14}N$ ratio measured in CN, is above the terrestrial value.

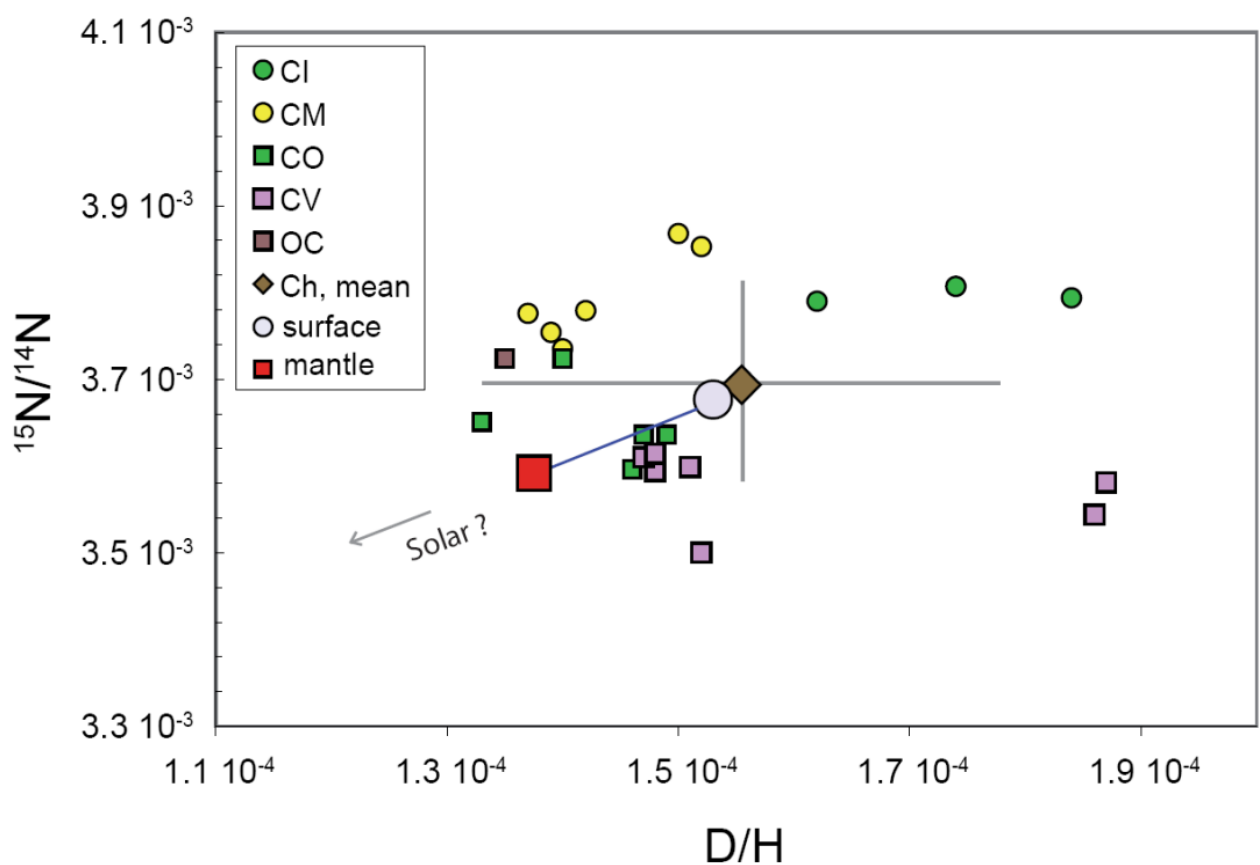

**Fig. 2** : Detail of Fig. 1. Two terrestrial values are given for the surface inventory ( large mauve dot, D/H and $^{15}N/^{14}N$ ratios for the oceans and the atmosphere, respectively) and for a putative mantle end-member (large square dot) defined with the lowest D/H ratio found in Hawaiian lava minerals (Deloule et al., 1991) and the lowest terrestrial $^{15}N/^{14}N$ ratio which was found in diamonds (Cartigny et al., 1997). The mean value of carbonaceous chondrites (excluding extreme values from CR-CB and the Bells CM chondrite) and ordinary chondrites, shown by the brown diamond, is very close to the terrestrial surface value. Enstatite chondrites are not represented because they do not contain sizeable amounts of water. The presumed mantle end-member composition is suggestive to the occurrence of a solar-like component at depth for N and H, as in the case of neon.

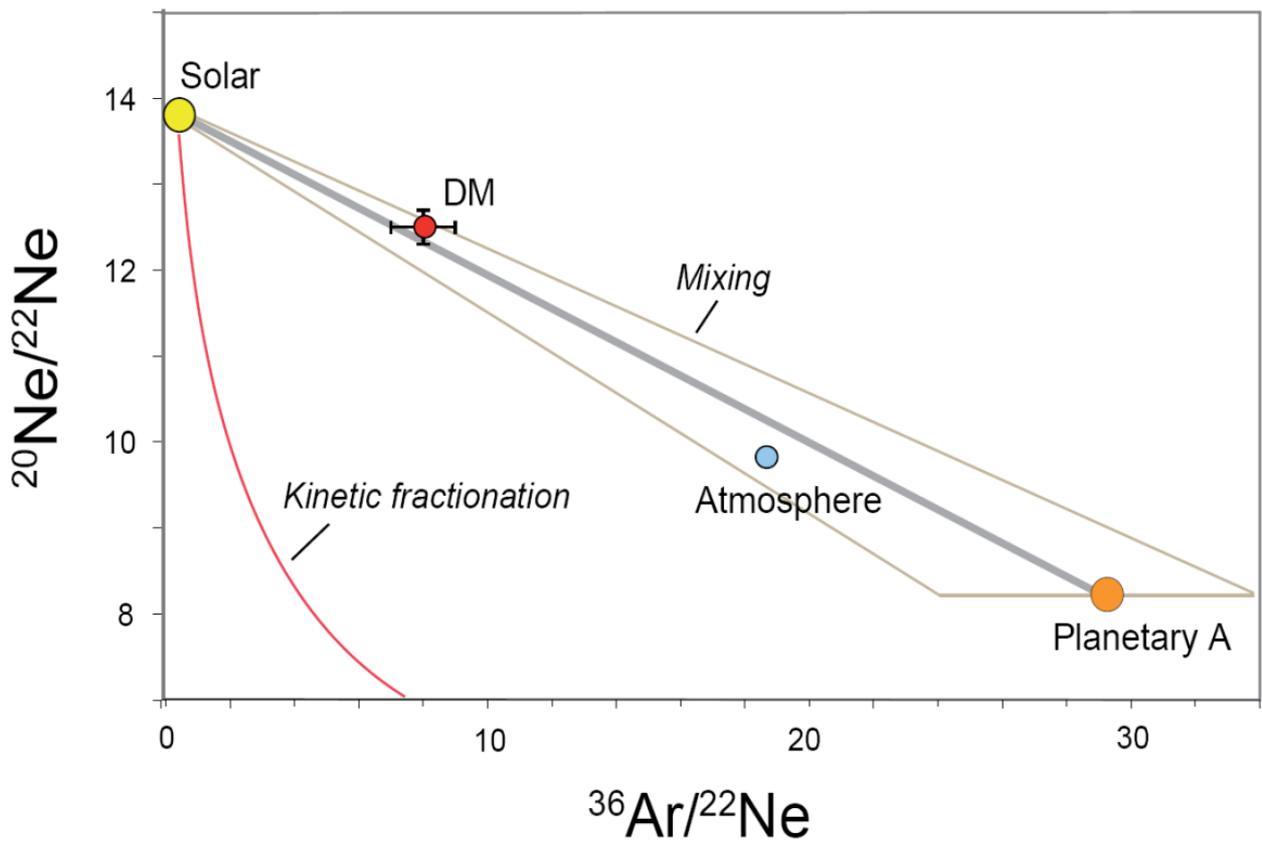

**Fig 3** : Mixing diagram between Solar and the Planetary-A component of Black (1971) and Mazor et al. (1970), defined by the lower boundary of the envelope delimiting carbonaceous chondrite data. "DM" is defined by the mean of the $^{36}Ar/^{22}Ne$ ratios of the Popping rock (Moreira et al., 1998) and the $CO_2$ well gases from Ballentine et al. (2005) on one hand, and the highest end-member $^{20}Ne/^{22}Ne$ ratio of MORBs (Trieloff et al., 2000). For comparison, the evolution of a solar composition upon kinetic fractionation (Rayleigh distillation, where the fractionation factor is the square root of masses) is represented by the curve at the left hand side. Neither the atmosphere nor the DM compositions fit this evolution trend.

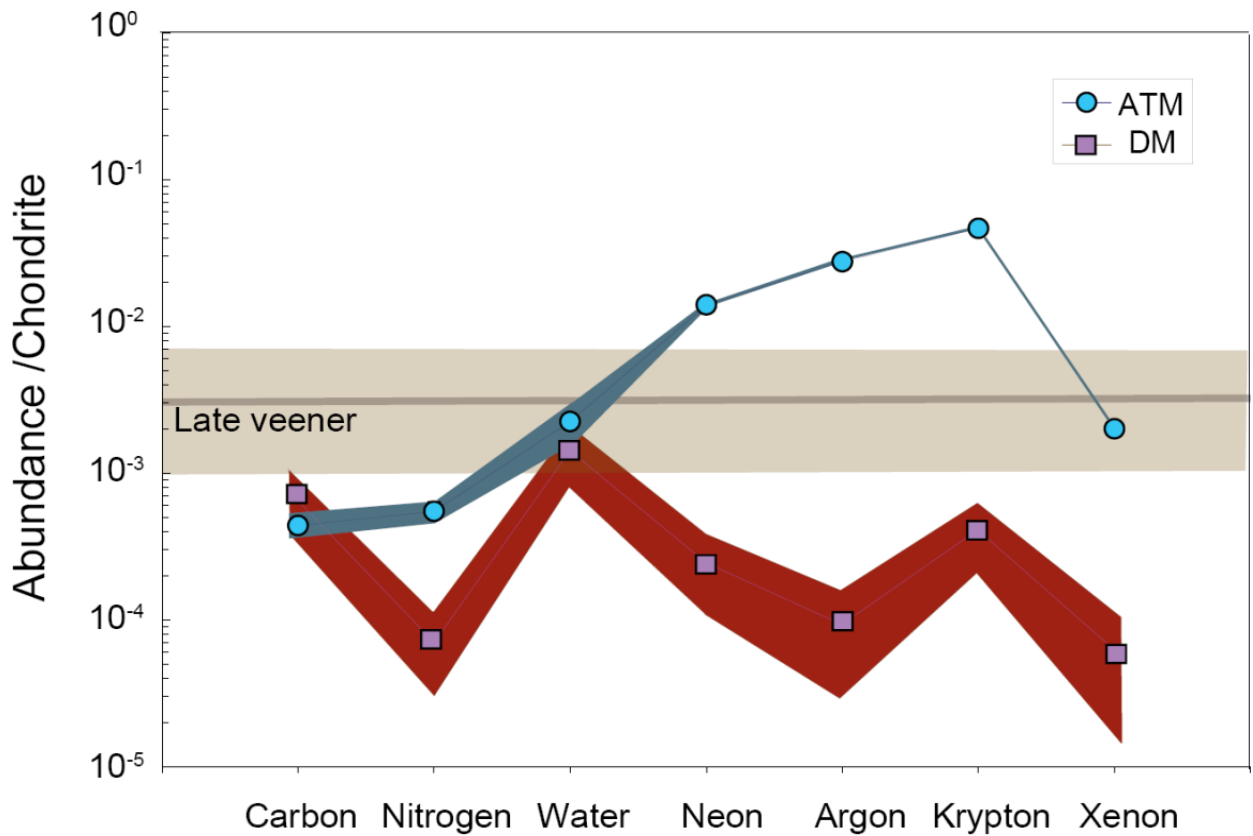

**Fig. 4** : Comparison of the abundance patterns of the atmosphere (surface inventory : air + oceans + sedimentary rocks) and of the depleted mantle (DM) . Abundances are normalized to the mass of the Earth for the atmosphere, and to the mass of the mantle for DM, and both to the carbonaceous chondrite composition. The shaded area represents the chondritic composition to account for the abundance of platinum group elements (PGEs) in the mantle (Righter, 2003).

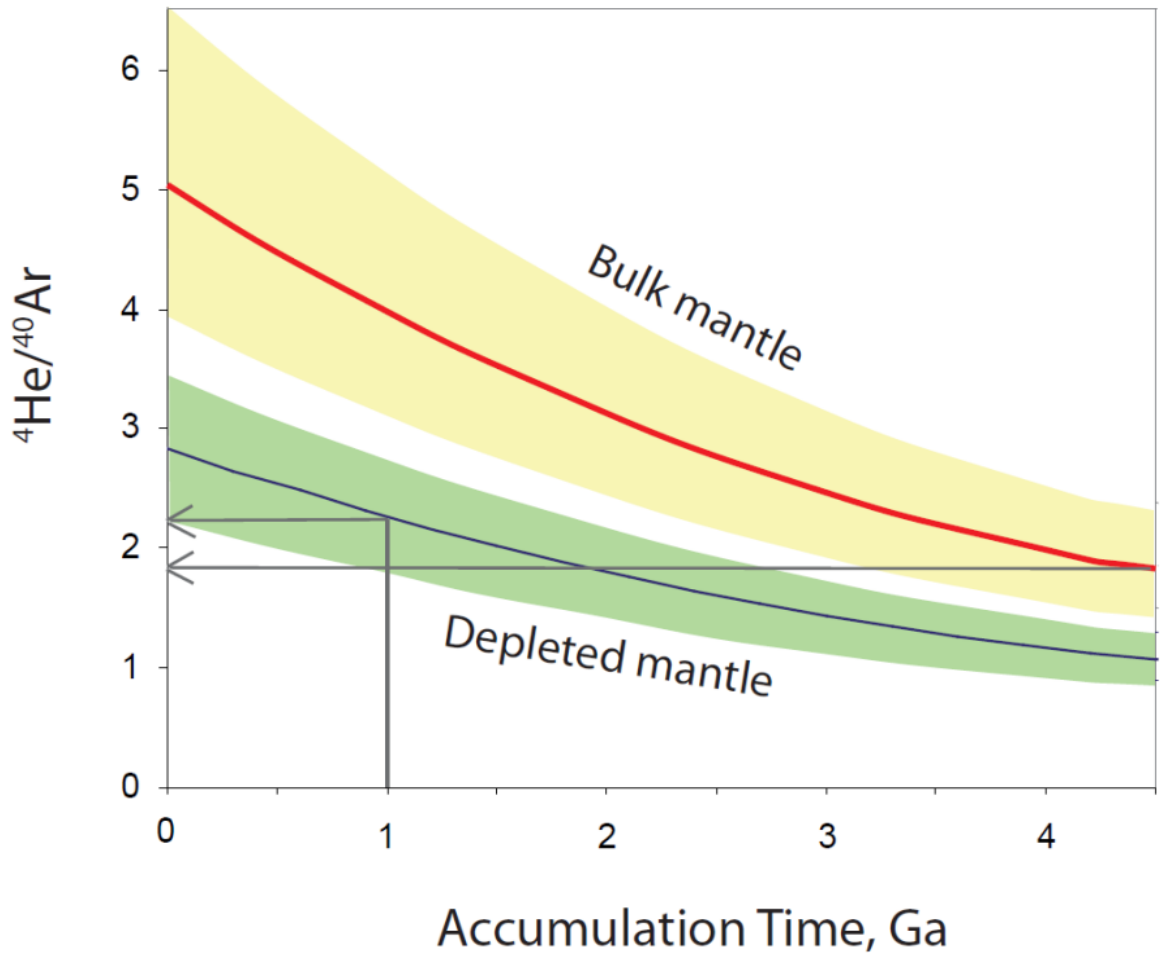

**Figure 5** : $^4$He/$^{40}$Ar evolution of mantle reservoirs as a function of time. Light grey area: OIB K/U and U/Th of 11900 (±2200) and 3.9 respectively; dark grey: MORB K/U and Th/U ratios of 19000 (±2600) and 3, respectively. The K/U ratios are from Arevalo et al. (2009) and the Th/U are from Vlastelic et al. (2006). Uncertainties of 30% have been attributed to the Th/U ratios although the uncertainties in Th/U do not significantly affect the range in $^4$He/$^{40}$Ar. The $^4$He/$^{40}$Ar values fro accumulation times of 4.5 Ga in the bulk mantle, and 1Ga in the depleted mantle, respectively, are shown by arrows.

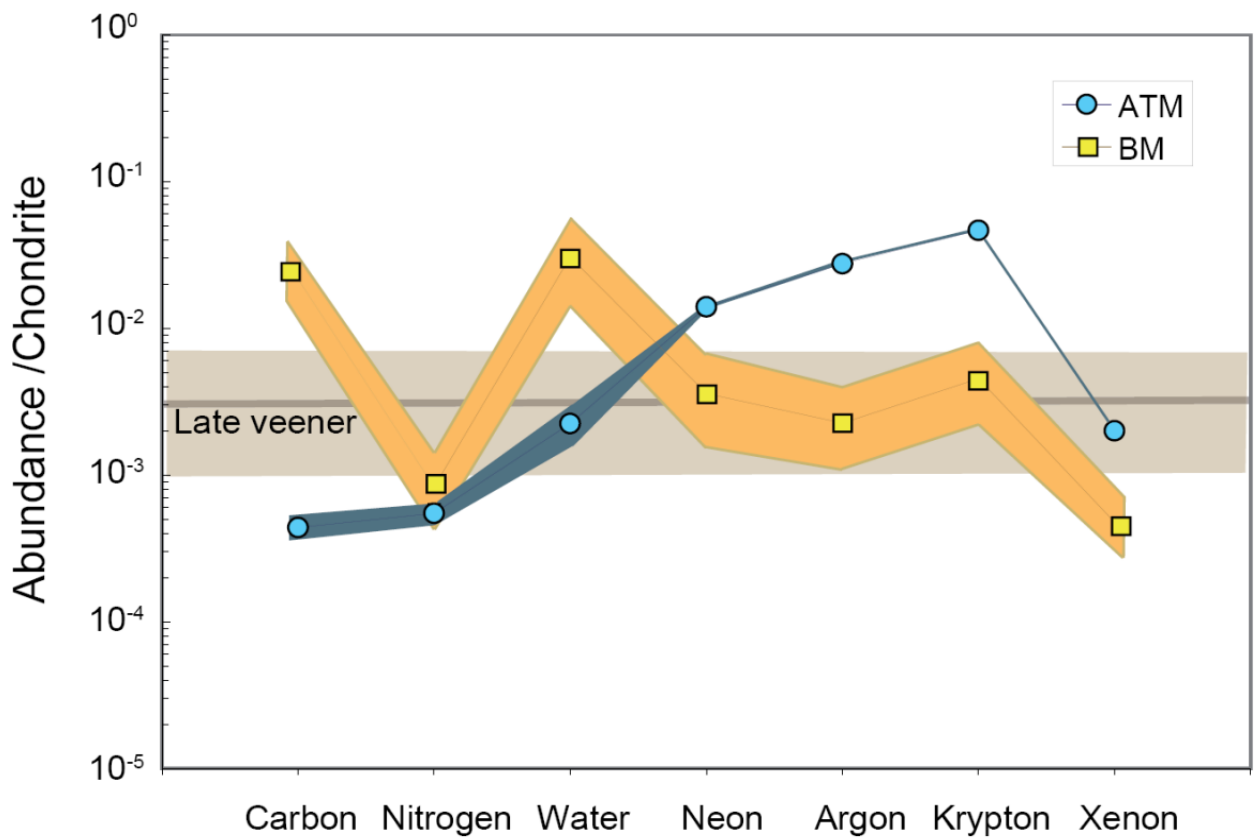

**Fig. 6** : Comparison of the abundance patterns of the atmosphere (ATM) and of the bulk mantle (BM). Same format as Fig. 4.

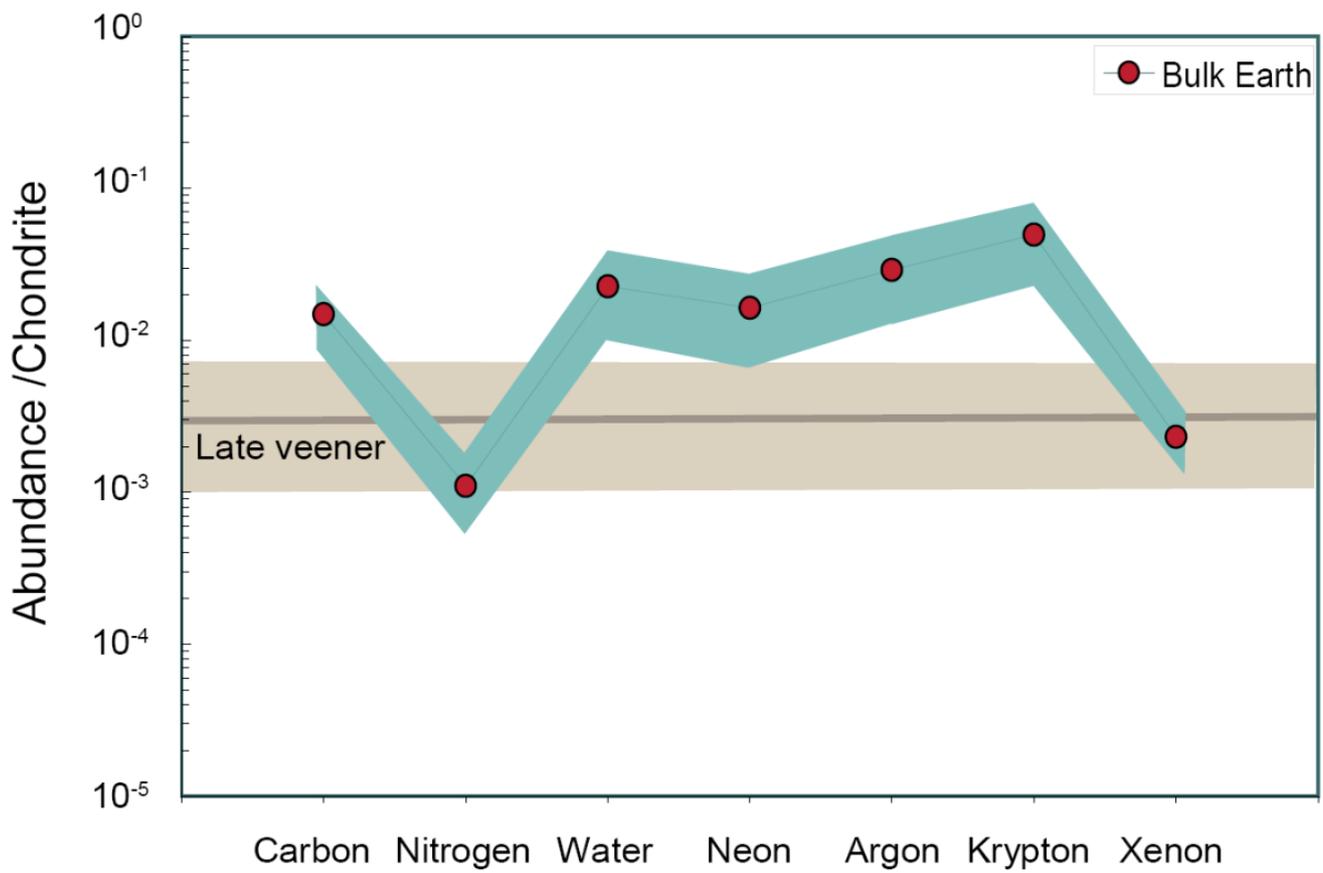

**Fig. 7** : Volatile composition of the bulk Earth, obtained by adding the number of atoms in the surface and the bulk mantle inventories, normalized to the mass of the Earth. Same format as Fig. 4.

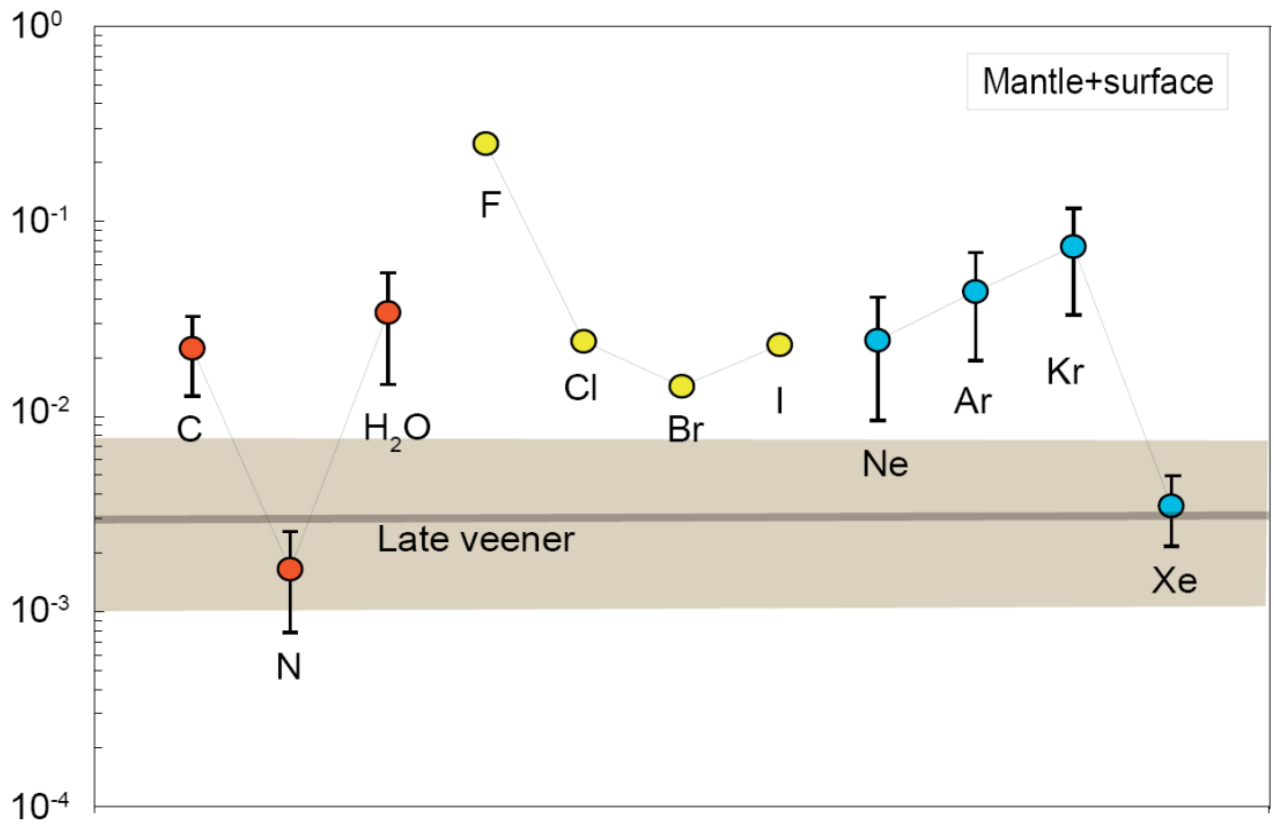

**Fig. 8** : Terrestrial composition of volatile elements extended to halogens. Data are from Deruelle et al. (1992) for iodine, and from compilation in McDonough and Sun (1995) for chondrites and bulk silicate Earth.

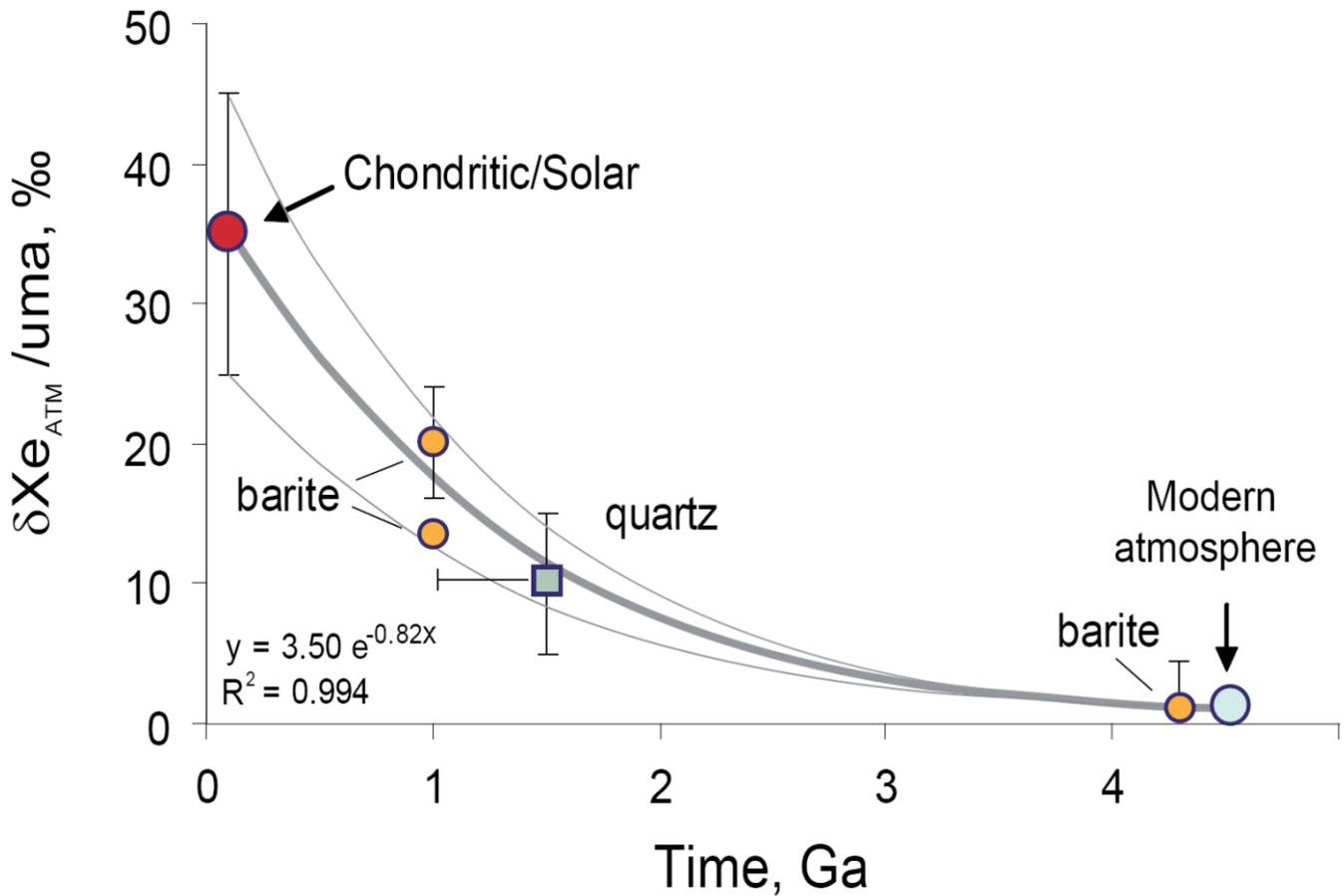

**Fig. 9** : Proposed evolution fof the Xe isotopic fractionation, expressed as a δ deviation, relative to modern atmospheric Xe, in per mil per mass unit, as a function of time (adapted from Pujol et al., 2011). The 3.5 Ga-old barite samples are from North Pole, Australia (Pujol et al., 2009; Srinivasan, 1976). The quartz sample data are from fluid inclusions trapped in hydrothermal quartz in 3.5-Ga old komatiite lava vacuoles. The Ar-Ar age of the fluid inclusions is 3.0 Ga. The younger, 170 Ma-old barite sample is from the Belorechenskoe deposit, Northern Caucasus, Russia (Meshik et al., 2001). The data fit an exponential decay curve, suggesting that the Xe isotopic fractionation was progressive with time.